# WiGig and IEEE 802.11ad    For Multi-Gigabyte-Per-Second WPAN and WLAN

SaiShankar N, Debashis Dash, Hassan El Madi, and Guru Gopalakrishnan
Tensorcom Inc., 5900 Pasteur Court, Carlsbad, CA 92008

**Abstract:**

The Wireless Gigabit Alliance (WiGig) and IEEE 802.11ad are developing a multigigabit wireless personal and local area network (WPAN/WLAN) specification in the 60 GHz millimeter wave band. Chipset manufacturers, original equipment manufacturers (OEMs), and telecom companies are also assisting in this development. 60 GHz millimeter wave transmission will scale the speed of WLANs and WPANs to 6.75 Gbit/s over distances less than 10 meters. This technology is the first of its kind and will eliminate the need for cable around personal computers, docking stations, and other consumer electronic devices. High-definition multimedia interface (HDMI), display port, USB 3.0, and peripheral component interconnect express (PCIe) 3.0 cables will all be eliminated. Fast downloads and uploads, wireless sync, and multi-gigabit-per-second WLANs will be possible over shorter distances. 60 GHz millimeter wave supports fast session transfer (FST) protocol, which makes it backward compatible with 5 GHz or 2.4 GHz WLAN so that end users experience the same range as in today's WLANs. IEEE 802.11ad specifies the physical (PHY) sublayer and medium access control (MAC) sublayer of the protocol stack. The MAC protocol is based on time-division multiple access (TDMA), and the PHY layer uses single carrier (SC) and orthogonal frequency division multiplexing (OFDM) to simultaneously enable low-power, high-performance applications.

**Keywords:** 60 GHz communications; IEEE standards; WiGig; 802.11ad; contention based access protocol; scheduled protocol; Beamforming; power save

## 1 Introduction

There is a huge amount of unlicensed spectrum available worldwide in the 60 GHz band. Academia and industry have turned to the 60 GHz spectrum because of the universal availability of unlicensed spectrum, the ever-growing number of user applications creating heavy data traffic, and the need to reduce data transfer times. Considerable efforts have been made to use this spectrum and spur the development of silicon, similar to what happened with the 2.4 GHz ISM band 15 years ago. 60 GHz millimeter wave technologies offers a way to provide end users with guaranteed quality of service (QoS) for different applications. Fig. 1 shows the allocation for 60 GHz in different countries [1].

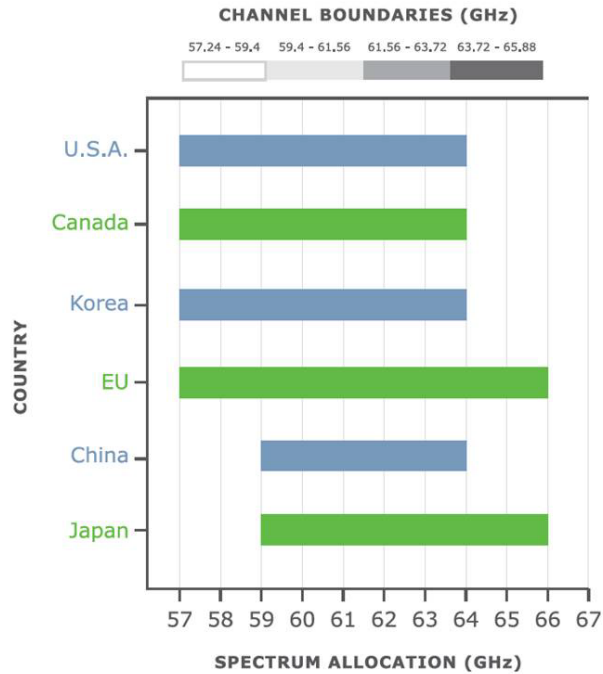

Figure 1. Spectrum allocation for WiGig.

60 GHz millimeter wave technologies create significant problems in designing the radio frequency (RF) front-end, processing gigabit-per-second data, and migrating to 40 nm and 28 nm low-power technologies in designing the silicon, considerable progress have been made in making it practical and feasible [4]-[6]. 60 GHz millimeter wave systems are needed to cater for newer applications, such as streaming video in the home or office, that have flourished as a result of last-mile access provided by internet service providers (ISPs). Such systems will also eliminate the need for cables around docking stations, and this will reduce clutter and allow easier connection between devices. There are multiple industry organizations involved in 60 GHz standardization, the notable ones being Wireless HD [7], IEEE 802.15.3c [8], WiGig [9], and IEEE 802.11ad [10]. The last two of these organizations involve a large number of silicon, OEM, and telecom companies that are motivated to have a single worldwide 60 GHz standard. WiGig began standardization in 2008 and has recently released the WiGig 1.0 standard. IEEE 802.11ad also began standardization in 2008 and has recently released IEEE 802.11ad Draft 9.0 standard. These standards are similar, and in this paper, we will refer to 802.11ad as the representative of both standards, pointing out when there is a feature that is unique to the WiGig standard. Similar standardization efforts have been made by ECMA-387 and CMMW Study Group [2], [3]. 60 GHz millimeter wave is the next wireless networking technology and will appear in the market around 2014 [11]. It is poised to repeat the successes of Bluetooth and Wi-Fi [12]. This explosive growth of the wireless industry in such a short time can also be attributed to the opening of unlicensed bands in 60 GHz by the Federal Communications Commission (FCC).

802.11ad aims to develop the protocol adaptation layers (PALs) to support a plethora of applications that will arise from the elimination of cables and from fast wireless sync and transfer. The PALs being considered by WiGig include wireless serial extension (WSE), which eliminates USB 3.0 cables; wireless bus extension (WBE), which eliminates PCIe 3.0 cables; wireless display extension (WDE), which eliminates high-definition multimedia interface (HDMI) and display port cables; and wireless secure digital (WSD), which makes secure digital input/output card (SDIO) disks wireless. The first important 60 GHz millimeter wave application to enter the market as wireless docking based on PCIe 3.0—with one second-generation lane (also called x2)—or USB 3.0. All devices with 802.11ad MAC/PHY/Radio use the corresponding PALs between the application and MAC layers to seamlessly transfer information between devices as if the devices were connected by wires. Another 60 GHz application is wireless HDMI based on WDE, which allows transfer of uncompressed bits from devices such as set top boxes and blue ray disc players to television screens and from laptops, desktops, or ultrabooks to monitors via a display port cable replacement. The WDE also supports H264 compressed rates for handling variations in the wireless channel and to ensure seamless content delivery to the end users. Performance of the PHY and MAC protocols is analyzed in [13] and [14].

In this paper, we describe the novel features of the MAC and PHY sublayers of the protocol stack defined in 802.11ad. In section 2, we describe the TDMA protocol and the need for directionality in 60 GHz. In section 3, we outline the 802.11ad PHY layer, and in section 4, we outline the MAC layer. In section 5, we outline the beamforming protocol, and in section 6 we outline the power-saving protocol. In section 7, we describe the fast session transfer, and in section 8, we show achievable rates using different MAC- and PHY-layer packet transmission options. Section 9 concludes the paper.

## 2 TDMA Protocol and the Need for Directionality

Interest in the 30–300 MHz millimeter wave spectrum has increased significantly because of low-cost, high-performance CMOS technology and because of low-loss, low-cost organic packaging. A millimeter-wave radio can be empowered for the same cost as a radio operating in the 5 GHz band or lower. This advantage, combined with wide available bandwidth, makes the millimeter-wave spectrum more attractive than ever before for supporting new systems and applications. A millimeter wave signal can propagate over a few kilometers at lower frequencies, penetrating through different construction materials and deriving advantages from reflection and refraction; however, they are highly directional and can be sustained only over short distances. The reason for this directionality is explained by the Friis free space equation:

$$P_R = P_T G_T G_R \lambda^2 / 4\pi R^2 \tag{1}$$

where $P_R$, $P_T$, $G_T$, $G_R$, $\lambda$ and $R$ is the receive power, transmit power, transmit antenna gain, receive antenna gain, wavelength, and distance between the transmitter and receiver, respectively. There is a 22 dB loss when we move from 5 GHz to 60 GHz. This loss is due to lower wavelength and can be offset by using directional antennas with higher gains. If 2 GHz bandwidth was used in 60 GHz and $P_T$ = 10 dBm, the noise figure $Nf$ = 10 dB, and the shadow fading margin $\sigma$ = 6 dB, 1 Gbit/s throughput could not be achieved (Fig. 2). Therefore, the gains of directional antennas must be exploited to achieve higher rates.

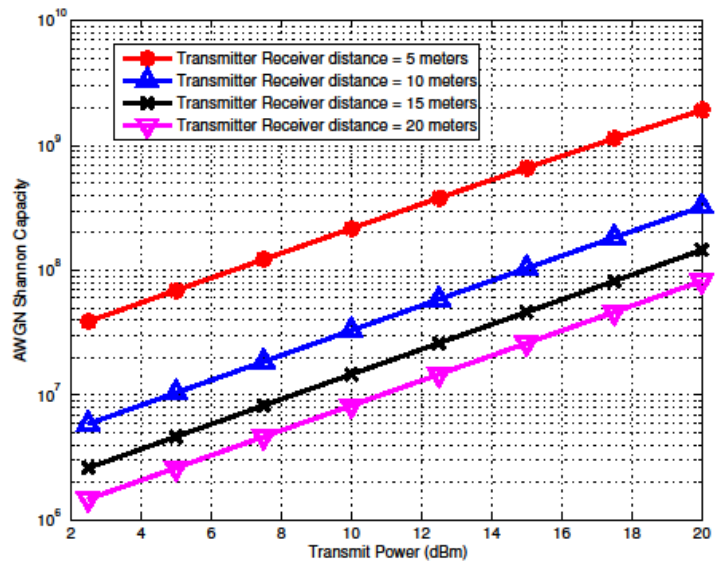

Figure 2. Shannon capacity versus transmit power at 60 Hz [1].

Directional communication requires complex discovery and beamforming protocols to establish links between different devices. Scheduled protocols such as TDMA are needed at the MAC layer to guarantee QoS at multi-gigabit-per-second rates. Randomized access protocols such as CSMA come with a random overhead that can depend on the number of users contending for the channel. Although CSMA/CA is still used to handle bursty traffic, allocation of contention-based access periods (CBAPs) is based on TDMA.

## 3 Physical Layer

802.11ad defines four different PHY layers: Control PHY, SC PHY, OFDM PHY and low-power SC PHY (LPSC PHY). Control PHY is MCS 0. SC starts at MCS 1 and ends at MCS 12; OFDM PHY starts at MCS 13 and ends at MCS 24; and LPSC starts at MCS 25 and ends at MCS 31. MCS 0 to MCS 4 are mandatory PHY MCSs. Here, we briefly describe the different PHYs and their packet structures. The system clock rate is 2640 MHz, and this rate is used for OFDM also. Control, SC and LPSC PHYs have a clock rate of 2/3 × 2640 = 1760 MHz.

### 3.1 General Packet Structure

As is common with all 802.11 packet formats, the packet consists of a short training sequence, a channel estimation sequence, the physical layer convergence procedure (PLCP) header, MAC packet, and cyclic redundancy check (CRC). Although there are different PHYs, they all have this unique structure, which ensures implementers do not need to change packet formats when using different PHYs. The only difference is that each PHY is a different size and uses a different Golay code.

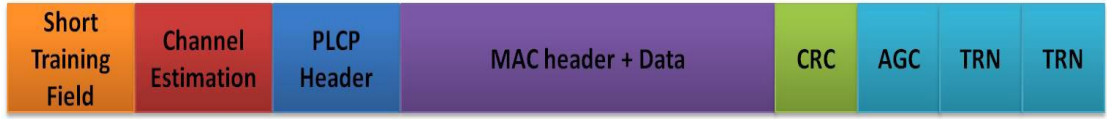

Figure 3. IEEE 802.11ad packet structure.

The short training field (STF) and channel estimation field (CEF) help signal acquisition, automatic gain control training, predicting the characteristics of the channel for the decoder, frequency offset estimation and synchronization. Both STF and CEF sequences use Golay codes. The PLCP header indicates the size of the packet as well as the modulation structure (MCS) of the packet. The MAC packet comprises MAC header and data and contains information about the destination. CRC ensures that the packet is not corrupted while being transmitted through the air. 802.11ad has a TRN field comprising Golay codes. This field is used in beam tracking and refinement and is described in section 5. Fig. 3 shows a typical packet in 802.11ad.

### 3.2 Control PHY

The control PHY, MCS 0, is the minimum rate that all devices use to communicate with before establishing a high-rate beamformed link. To help discovery and detection, the control PHY has an STF comprising 50 Golay sequences, each of which is 128 samples long. The CEF that follows the STF has nine Golay sequences. The STF comprises 48 × $Gb128(n)$ (each 128 samples long) followed by a single repetition of -$Gb128(n)$. The CEF comprises $Gu512(n)$ and $Gv512(n)$ followed by $Gv128(n)$. These sequences are represented as function of $Ga128(n)$ and $Gb128(n)$ in section 21.11 of the IEEE 802.11ad draft specification [10]. $Gu512(n)$ and $Gv512(n)$ are given by:

$$Gu512(n) = [-Gb128 -Ga128 \, Gb128 \, Ga128] \qquad (2)$$

$$Gv512(n) = [-Gb128 \, Ga128 -Gb128 -Ga128] \qquad (3)$$

The control PHY uses BPSK with a code rate of 1/2 and is spread using a 32 code to create a PHY rate of 27.5 Mbit/s. The Control PHY is used for transmitting and receiving frames such as beacons, information request and response, probe request and response, sector sweep, sector sweep feedback, and other management and control frames. It provides reliability and exploits gain of the transmit antenna. Additionally, the first frame transmitted during the beam refinement protocol (BRP) phase is also a control PHY frame.

### 3.3 Single-Carrier PHY, OFDM PHY and LPSC PHY

MCSs 1 to 4 are mandatory and ensure that all devices, irrespective of their PHY, are interoperable. All the MCSs, with the exception of LPSC PHY, use LDPC code, and the LPSC uses Reed Solomon (RS) codes. The following two subsections describe the MAC header and data packet encoding process for SC PHY. Other PHYs use a similar encoding process. All packets are modulated using BPSK, QPSK, 16-QAM and 64-QAM (Table 1).

Table 1. PHY modulation and coding scheme table

| MCS Index | Modulation | $N_{CBPS}$ | Repetitions | Code Rate | $N_{BPCS}$ | $N_{DBPS}$/Coding | DataRate (Mbit/s) |
|---|---|---|---|---|---|---|---|
| 0 | π/2 BPSK | 1 | 32 | 1/2 | 1 | 168 | 27.5 |
| 1 | π/2 BPSK | 1 | 2 | 1/2 | 1 | 168 | 385.0 |
| 2 | π/2 BPSK | 1 | 1 | 1/2 | 1 | 168 | 770.0 |
| 3 | π/2 BPSK | 1 | 1 | 5/8 | 1 | 168 | 962.5 |
| 4 | π/2 BPSK | 1 | 1 | 3/4 | 1 | 168 | 1155.0 |
| 5 | π/2 BPSK | 1 | 1 | 13/16 | 1 | 168 | 1251.25 |
| 6 | π/2 QPSK | 2 | 1 | 1/2 | 1 | 168 | 1540.0 |
| 7 | π/2 QPSK | 2 | 1 | 5/8 | 1 | 168 | 1925.0 |
| 8 | π/2 QPSK | 2 | 1 | 3/4 | 1 | 168 | 2310.0 |
| 9 | π/2 QPSK | 2 | 1 | 13/16 | 1 | 168 | 2502.5 |
| 10 | π/2 16 QAM | 4 | 1 | 1/2 | 1 | 168 | 3080.0 |
| 11 | π/2 16 QAM | 4 | 1 | 5/8 | 1 | 168 | 3850.0 |
| 12 | π/2 16 QAM | 4 | 1 | 3/4 | 1 | 168 | 4620.0 |
| 13 | SQPSK | 336 | 1 | 1/2 | 1 | 168 | 693.0 |
| 14 | SQPSK | 336 | 1 | 5/8 | 1 | 210 | 866.25 |
| 15 | QPSK | 672 | 1 | 1/2 | 2 | 336 | 1386.0 |
| 16 | QPSK | 672 | 1 | 5/8 | 2 | 420 | 1732.5 |
| 17 | QPSK | 672 | 1 | 3/4 | 2 | 504 | 2079.0 |
| 18 | 16-QAM | 1344 | 1 | 1/2 | 4 | 672 | 2772.0 |
| 19 | 16-QAM | 1344 | 1 | 5/8 | 4 | 840 | 3465.0 |
| 20 | 16-QAM | 1344 | 1 | 3/4 | 4 | 1008 | 4158.0 |
| 21 | 16-QAM | 1344 | 1 | 13/16 | 4 | 1092 | 4504.0 |
| 22 | 64-QAM | 2016 | 1 | 5/8 | 6 | 1260 | 5179.0 |
| 23 | 64-QAM | 2016 | 1 | 3/4 | 6 | 1512 | 6237.0 |
| 24 | 64-QAM | 2016 | 1 | 13/16 | 6 | 1638 | 6756.75 |
| 25 | π/2 BPSK | 392 | 1 | 13/16 | 6 | RS(224,208)+BS(16,8) | 626.0 |
| 26 | π/2 BPSK | 392 | 1 | 13/16 | 6 | RS(224,208)+BS(12,8) | 834.0 |
| 27 | π/2 BPSK | 392 | 1 | 13/16 | 6 | RS(224,208)+SPC(9,8) | 1112.0 |
| 28 | π/2 QPSK | 392 | 1 | 13/16 | 6 | RS(224,208)+BS(16,8) | 1251.0 |
| 29 | π/2 QPSK | 392 | 1 | 13/16 | 6 | RS(224,208)+BS(12,8) | 1668.0 |
| 30 | π/2 QPSK | 392 | 1 | 13/16 | 6 | RS(224,208)+SPC(9,8) | 2224.0 |

| 31 | π/2 QPSK | 392 | 1 | 13/16 | 6 | RS(224,208)+ BC(8,8) | 2503.0 |

## 3.4 Header Encoding

The header is encoded using a single SC block of $N_{CBPB}$ symbols with $N_{GI}$ guard symbols. The bits are scrambled and encoded in the following steps:

1) The input header bits $(b_1, b_2,...,b_{LH})$ LH = 64 are scrambled, starting from the eighth bit, in order to create $d1s = (q_1, q_2,...,q_{LH})$.
2) The LDPC code word $c = (q_1, q_2,...,q_{LH}, 0_1, 0_2,...,0_{504-LH}, p_1, p_2,...,p_{168})$ is created by concatenating 504-LH zeros to the LH bits of d1s and then generating the parity bits $p_1, p_2,...,p_{168}$ so that $Hc^T = 0$, where H is the parity-check matrix for the 3/4 LDPC code specification in IEEE 802.11ad.
3) Bits LH + 1 through 504 and bits 665 through 672 of the code word c are removed to create the sequence $cs1 = (q_1, q_2,...,q_{LH}, p_1, p_2,...,p_{160})$.
4) Bits LH +1 through 504 and bits 657 through 664 of the code word c are removed to create the sequence $cs2 = (q_1, q_2,...,q_{LH}, p_1, p_2,...,p_{152}, p_{161}, p_{162},...,p_{168})$ and then to create XOR with a PN sequence. The PN sequence is generated from the LFSR used for data scrambling, and the LFSR is initialized to the all-ones vector.
5) $cs_1$ and $cs_2$ are concatenated to form the sequence $(cs_1, cs_2)$. The resulting 448 bits are then mapped as π/2-BPSK, and the NGI guard symbols are prepended to the resulting NCBPB symbols.

## 3.3 Data Encoding

The data packet is encoded using LDPC, which includes deciding the number of shortening/repetition bits in every code word, shortening, coding each word, and repetition of bits. Data packet encoding occurs in the following steps:

1) The number of LDPC code words is given by $N_{CW} = (length \times 8 \rho)/(L_{CW} \times R)$. This is used to calculate the number of datapad bits given by $NDATA\_pad = (N_{CW} \times L_{CW} \times R)/\rho - (length \times 8)$, where $L_{CW} = 672$ is the LDPC code word length; length is the length of the PSDU defined in the header field (in octets); ρ is the repetition factor (1 or 2); and R is the code rate. The scrambled PSDU is concatenated with $NDATA_P$ AD zeros, which are scrambled using the continuation of the sequence that scrambled the PSDU input bits.

2) The output stream of the scrambler is broken into blocks of $L_{CWD} = L_{CW} \times R$ bits so that the mth data word is $b_1^m, b_2^m,..., b_{L_{CWD}}^m \quad m < N_{CW}$

3) To each data word, n - k = $L_{CW} - (R \times L_{CW})$ parity bits $p_1^m, p_2^m,..., p_{n-k}^m$ are added to create the code word $c^m = b_1^m, b_2^m,..., b_{L_{CWD}}^m, p_1^m, p_2^m,..., p_{n-k}^m$ so that $H_c^{(m^T)} = 0$

4) The code words are concatenated one after the other to create the coded bitstream $c_1, c_2,..., c_{L_{CWD} \times N_{CW}}$ The number of symbol blocks is given by $N_{BLK} = (N_{CW} \times L_{CW}) / N_{CBPB}$, and the number of symbol block padding bits is given by $N_{BLKPAD} = (N_{BLK} \times N_{CBPB}) - (N_{CW} \times L_{CW})$, where $N_{CBPB}$ is the number of coded bits per symbol block.

5) The coded bitstream is concatenated with $N_{BLKPAD}$ zeros, which are scrambled using the continuation of the scrambler sequence that scrambled the PSDU input bits. Table I shows the MCS values allowed in the IEEE 802.11ad specification.

## 4 Overview of 802.11AD MAC Protocol

### 4.1 802.11ad Superframe

The 802.11ad superframe is called the beacon interval and comprises a beacon transmission interval (BTI), a data transfer interval (DTI), and optional association beamforming training (A-BFT) or announcement transmission intervals (ATI) (Figs. 4 and 10). The DTI can include one or more service periods (SPs) and contention-based access periods (CBAPs).

### 4.2 Service Period Channel Access

An SP is a scheduled access period between two stations: a transmitter and receiver. SPs are suitable for directional (high-gain) antenna use. New features introduced by 802.11ad include spatial sharing, where SPs need not be out of synch, that is, an SP between stations A and B may occur at the same time as another SP between stations C and D. IEEE 802.11ad also introduces dynamic SP allocation, truncation, and extension. An SP allocation is dynamic if it is not initially scheduled by the personal basic service set (PBSS) control point/access point (PCP/AP) but is scheduled during an existing SP or CBAP. SP truncation occurs when the transmitter station relinquishes the remaining time in its SP. SP extension occurs when the transmitter station extends the SP duration it had been allocated.

### 4.3 CBAP Channel Access

During a CBAP, all stations contend for channel access using a hybrid TDMA-CSMA/CA scheme based on 802.11 enhanced distributed channel access (EDCA). 802.11ad provides physical carrier sensing mechanism, provided by the physical layer, and a virtual carrier-sensing mechanism, provided by the MAC layer. Physical carrier sensing uses clear channel assessment (CCA), and

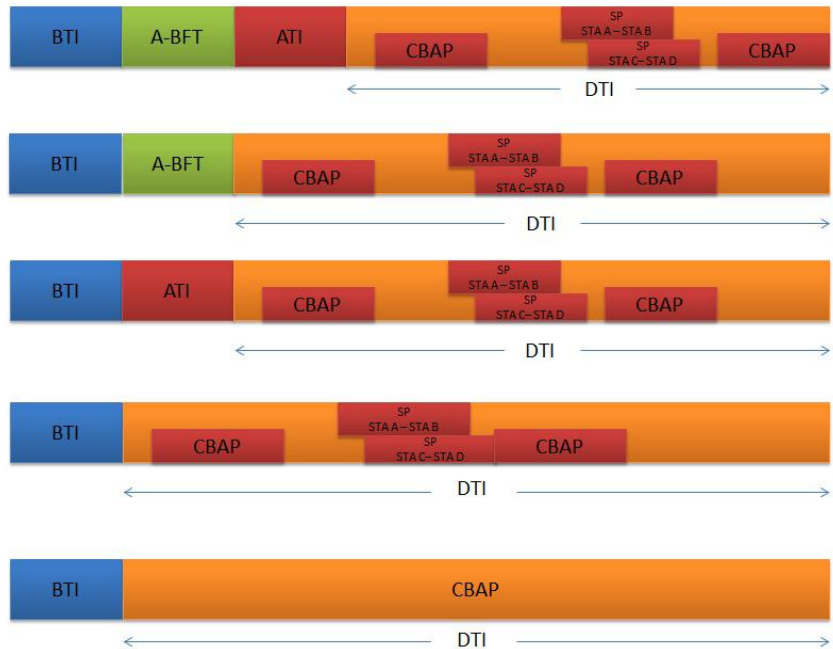

BTI: beacon transmission interval
DTI: data transfer interval
CBAP: contention-based access periods
ATI: announcement transmission intervals
STA: peer station

Figure 4. Examples of 802.11ad beacon intervals.

Virtual carrier sensing uses a timer called network allocation vector (NAV). NAV indicates, in microseconds, how long the channel is reserved by another station and counts down to 0. The virtual carrier-sensing mechanism uses request-to-send/directional multigigabit clear-to-send (RTS/DMG CTS) frames. When a station receives an RTS/DMG CTS frame, it sets its NAV to the value in the Duration field in the MAC header of the RTS/DMG CTS. Stations also use the Duration field of other frames to update their NAVs; however, the frame's destination address must be different from the receiving station's MAC address, and the value of the Duration field must be greater than the current NAV value. Stations may have a unique NAV or may have one NAV per sector. If a station has multiple NAVs and at least one has a non-zero value, the virtual carrier-sensing mechanism considers the medium busy. The medium is considered busy if either the physical or virtual carrier-sensing mechanism indicates it is busy; otherwise, it is considered idle. 802.11ad defines four different access categories (ACs) that have different priorities based on the user priority (UP) of the data being transferred. In order of increasing priority, these ACs are background (BK), best effort (BE), video (VI), and voice (VO). Only BE is mandatory in the standard, that is, only BE is implemented or all four are. When all four ACs are implemented, all ACs within a given station have to contend with each other and with other stations for channel access. Each AC contends for a channel in the following way: After the medium has been idle for a period of time, called the arbitration interframe space (AIFS [AC]), the station contending for access randomly sets its backoff timer to a between 0 and the contention window (CW [AC]).

CW [AC] is initialized to $CW_{min}$ [AC] and is updated after every transmission. In case of transmission failure, CW [AC] is updated using

$$CW\,[AC] = 2 \times CW\,[AC] + 1 \qquad (4)$$

When CW [AC] reaches $CW_{max}$ [AC], it remains unchanged for any remaining retries. In the case of transmission success, CW [AC] is reset to $CW_{min}$ [AC]. At every slot time boundary, the medium is sensed. If the medium is found to be idle, the backoff timer is decremented; otherwise, it is suspended. When the backoff timer for a particular AC reaches 0, that AC obtains exclusive channel access for a period of time called the transmit opportunity (TXOP [AC]). During the TXOP [AC], only frames with UP mapping to that AC may be transmitted. If the backoff timers of two or more ACs reach zero at the same time, channel access is granted to the AC with the highest priority, and the other ACs treat this occurrence as if it were an external collision that happened in the wireless medium. The other ACs then enter backoff phase. For each AC, EDCA parameters such as $CW_{min}$ [AC], $CW_{max}$ [AC], AIFS [AC] and TXOP [AC] are calculated by the PCP/AP and included in the DMG beacon, probe response, or (re)association response frames transmitted by the PCP/AP. Higher-priority ACs are granted lower values for $CW_{min}$ [AC], $CW_{max}$ [AC], AIFS [AC] so that they can gain channel access while lower-priority ACs are still in backoff phase.

**4.4 Packet Aggregation**
802.11ad is capable of a multi-gigabyte-per-second data rate because of features such as packet aggregation and block acknowledgement in addition to directional antennas. The basic MAC data unit is called MAC protocol data unit (MPDU) and comprises a MAC header and a MAC service data unit (MSDU) or MAC payload. A PHY header and an MPDU comprise a PHY PDU (PPDU). 802.11ad uses the Galois/counter mode (GCM) protocol for data encryption. This protocol was designed for encryption at multi-gigabytes-per-second data rates.

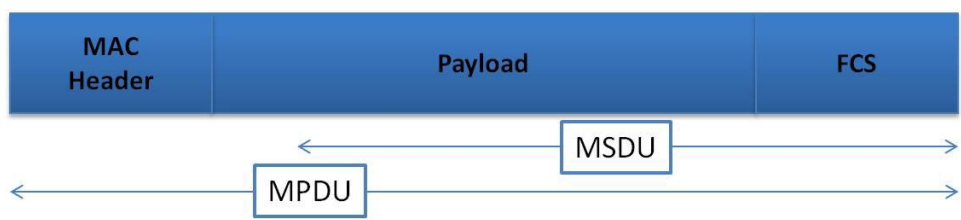

MPDU: MAC protocol data unit
MSDU: MAC service data unit

Figure 5. 802.11ad MPDU structure.

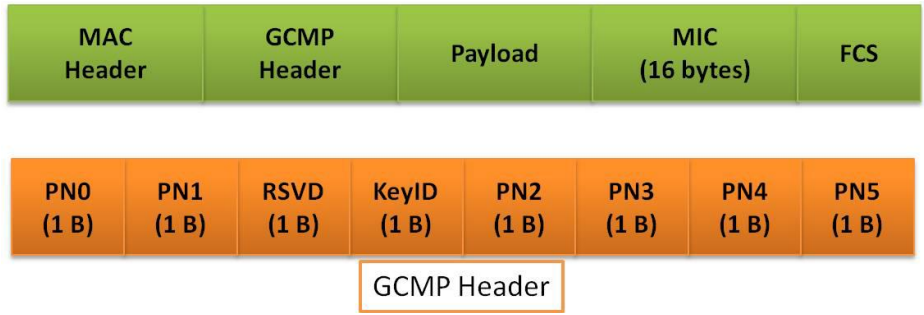

GCMP: GCM protocol

Figure 6. 802.11ad MPDU structure with encryption turned on.

An encrypted MPDU includes a GCM protocol header and a MIC field. Fig. 5 shows the MPDU structure. Fig. 6 shows the MPDU structure with encryption turned on. Packet aggregation involves combining several packets into a single packet. When several MSDUs or MPDUs are combined, the resulting packet is called aggregated MSDU (A-MSDU) (Fig. 7) or aggregated MPDU (A-MPDU) (Fig. 8). 802.11ad uses a new type of packet aggregation called aggregated PPDU (A-PPDU). In an A-PPDU packet, several PPDUs are transmitted back-to-back without interframe spacing (IFS) and preamble in between. A-PPDU reduces overhead associated with IFS and MAC/PHY header processing.

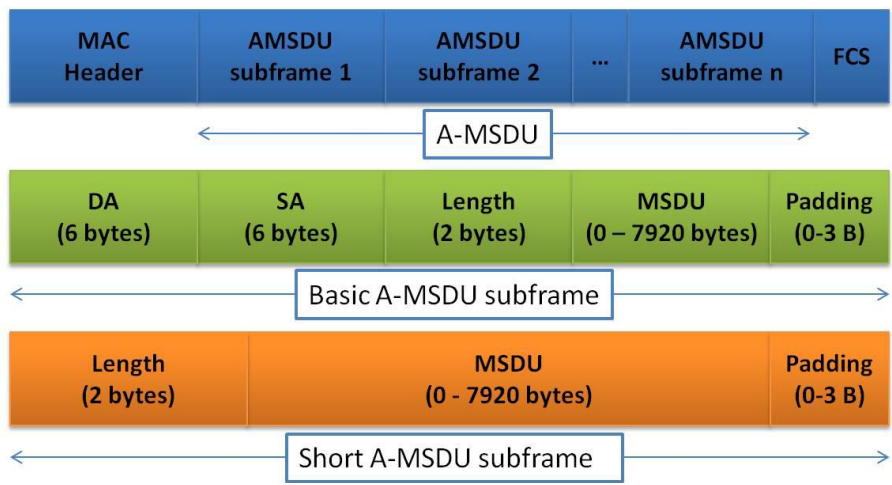

AMSDU: aggregated MAC service data unit
FCS: frame check sequence

Figure 7. 802.11ad A-MSDU structure.

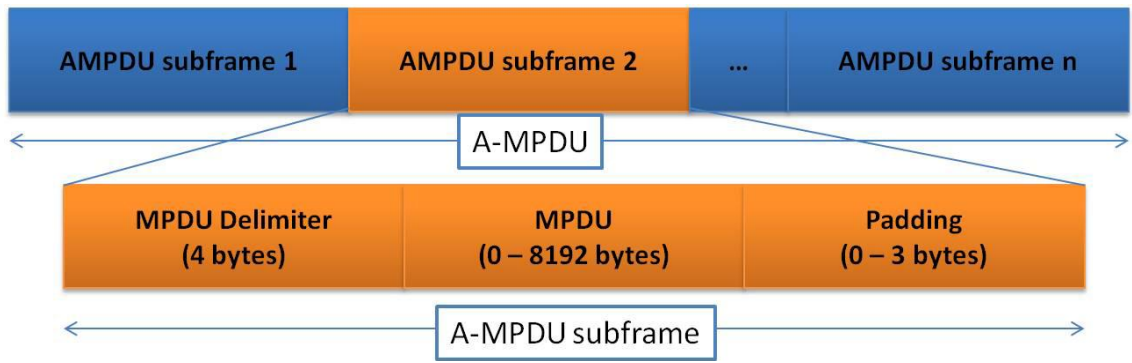

AMPDU: aggregated MAC protocol data unit

Figure 8. 802.11ad A-MPDU structure.

### 4.5 Acknowledgement Policies
802.11ad defines a frame acknowledgement (ACK) policy called block acknowledgement. When block ACK is enabled, the transmitting station transmits a block of frames one frame at a time immediately after each other without waiting for the receiver to acknowledge the previous frame. After the entire block of frames has been transmitted, the receiving station sends a control frame called block ACK that includes a bitmap. The bitmap, in which each bit corresponds to a frame, indicates which frames were received successfully and which ones were not. The receiver knows to send a block ACK frame when it receives a block ACK request frame from the transmitter. This ACK policy allows the transmitter to use shorter IFS between frames, and it eliminates the IFS between each frame and its individual ACK frame, as in a typical stop-and-wait protocol. Fig. 9 shows normal ACK and block ACK policies.

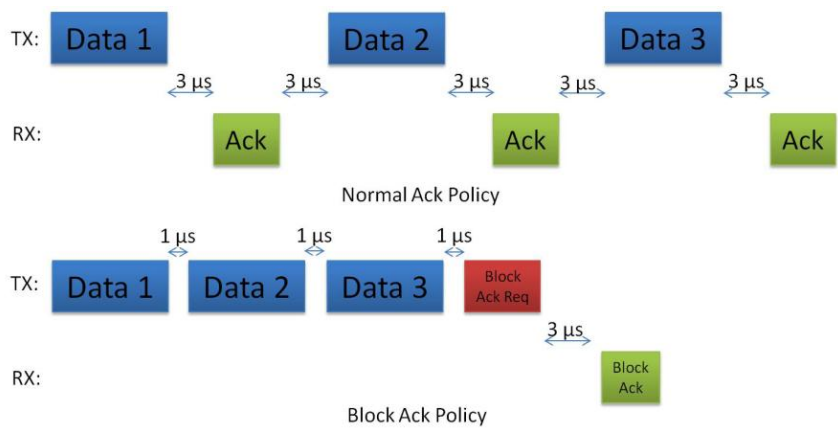

**Figure 9.** Normal ACK policy and block ACK policy.

### 5 Beamforming Protocol
Because of the highly directional nature of 60 GHz communications, the transmitter and receiver antennas need to be aligned in the right direction to obtain maximum gain. 802.11ad supports up to four transmitter antennas, four receiver antennas, and 128 sectors.

Beamforming is mandatory in 802.11ad, and both transmitter-side and receiver-side beamforming are supported.

Beamforming can be done at the transmitter side, receiver side, or at both sides [15]. Transmitter-side beamforming usually requires feedback from the receiver, especially when the transmitter-to-receiver and receiver-to-transmitter channels are not reciprocal. The need for feedback can be reduced or eliminated by using space time codes; however, this can cause considerable overhead in the setting-up beamforming [16]. 802.11ad uses a selection-based protocol in which the transmitter sends training from certain sectors that are pre-defined according to distinct antenna patterns created by changing the antenna weights [17]. The receiver antenna maintains an omnidirectional pattern and measures the strength of the received signal from the different sectors. It responds with information about the best sector and measured quality. With this feedback, the transmitter chooses the best sector to use while transmitting to the receiver. Similarly, in receiver-side training, the receiver repeats the training from the transmitter, which is sent using an omnidirectional antenna pattern, and measuring the strength of the received signal through pre-defined receive sectors.

The station that starts the beamforming training is called the initiator, and the recipient is called the responder. Beamforming in 802.11ad involves sector-level sweep (SLS), beam refinement protocol (BRP), and beam tracking (BT). Fig. 10 shows the sequence of this beamforming. Each of the steps in the sequence is allowed in a particular part of the beacon interval (Fig. 11). SLS enables reliable communication at the lowest supported rate (called MCS0 in 802.11ad). Usually, transmitter-side training is done during the SLS. The BRP enables receiver training and iteratively trains the transmitter and receiver sides to improve on the values found during the SLS. Both SLS and BRP phases use their own special packets for beamforming training. By contrast, BT can be done during data transmission. It is used to track the beamforming state and improve it during data transmission. BT is implemented by adding training (TRN) fields to the back of a data packet.

The SLS involves initiator sector sweep (ISS), responder sector sweep (RSS), sector sweep feedback (SSW-FB), and sector sweep acknowledgement (SSW-ACK). The BRP comprises setup, multiple sector ID detection (MID), beam combining (BC), and BRP transactions. Of these, MID and BC are optional features for 802.11ad supporting stations. BT comprises BT request and BT response. The parameters for exchanging beamforming packets are obtained using the capability element in the beacon packets, probe request/response packets, or information request/response packets.

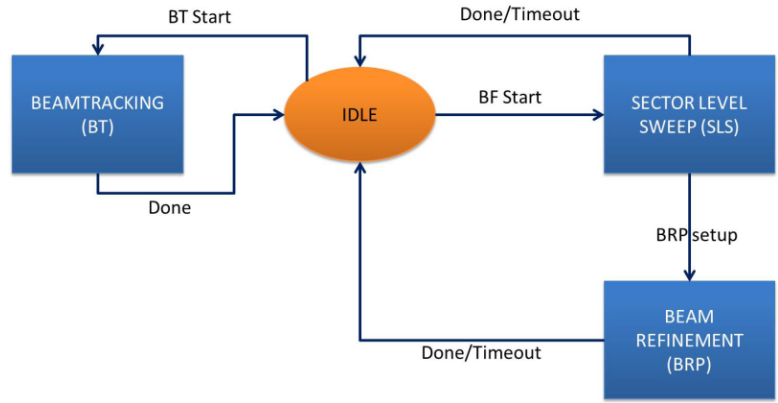

Figure 10. The sequence in which the different phases of beamforming occur in 802.11ad/WiGig standard.

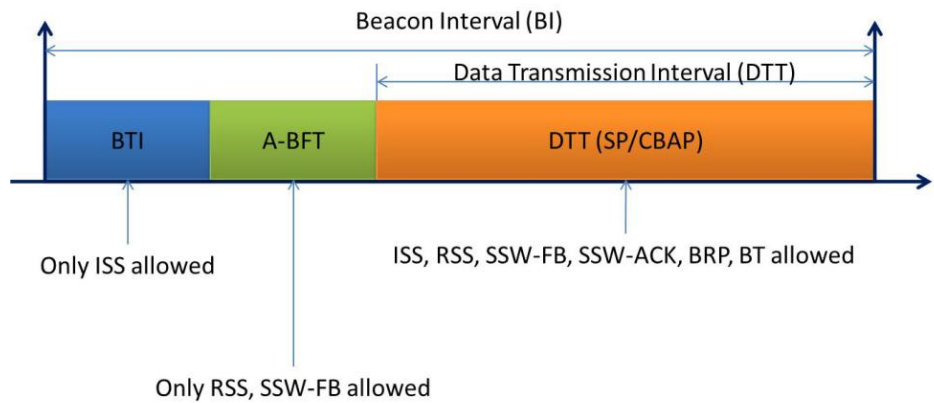

Figure 11. Different transmission periods and BF phases allowed in each part of the beacon interval.

## 5.1 Beaconing and Sector-Level Sweep

At the start of every BTI, the PCP/AP MAC schedules an initiator transmit sector sweep (TXSS) to transmit beacons through all sectors. The PCP/AP can also fragment the TXSS across multiple beacon intervals if the BTI is insufficient and cannot complete the TXSS. A station without a PCP/AP uses an omnidirectional receiving antenna configuration to scan non-associated beacons or receive associated beacons from the PCP/AP and determine the best sector/antenna ID using the sector sweep field at the end of TXSS. In a beacon, CDOWN is the number of pending beacon transmissions for completion of TXSS, with 0 being completion.

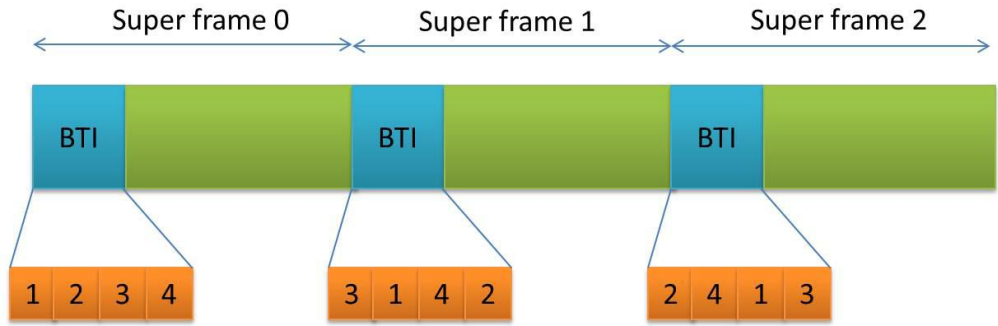

**Figure 12.** A sample DMG beacon transmission by a PCP having one transmit antenna with four sectors.

If multiple transmit antennas are supported, a PCP/AP station cannot switch its transmit antennas for beacon transmission within a BTI nor can it transmit a beacon more than once using the same antenna configuration. To minimize potential interference, the PCP/AP changes the order of sectors across beacon intervals if multiple directional beacon transmissions are required or waits for a random delay at the start of beacon interval if only a single beacon is to be transmitted (Fig. 12).

### 5.2 A-BFT Protocol

The PCP/AP announces the existence of an A-BFT period in the beacons, and this information is used for association and beamforming training for new stations, such as PBSS stations, that join the network. A-BFT may not be present in each beacon interval and may be periodically inserted by the AP/PCP. The A-BFT period allows stations to perform RSS and SSW-FB phases of beamforming with the PCP/AP. It is assumed that the new station has already used the beacons sent from all the transmit sectors of the PCP/AP to perform an ISS with the PCP/AP during the BTI. The A-BFT period is a slotted phase where each slot is a multiple of the time required for RSS and SSW-FB. The new stations use random backoff to select the A-BFT slot for an RSS. When beamforming is done during A-BFT, the SSW-ACK phase is skipped, and BRP is done during the data transmission interval if necessary. There may be no chance to do RSS during A-BFT because stations use random access; therefore, the AP/PCP may schedule an SP to continue beamforming with the particular station.

### 5.3 Sector-Level Sweep

The SLS is the basic type of beamforming supported by 802.11ad. It comprises ISS, RSS, SSW-FB and SSW-ACK. The link from the initiator to the responder is called the initiator link, and the link from the responder to the initiator is called the responder link. During SLS, the ISS phase is used to train the initiator link, and the RSS is used to train the responder link. The RSS contains feedback about the best sector found during ISS, and the SSW-FB contains the best sector found in the RSS. In SLS is concluded with an SSW-ACK (Fig. 13). A station can have separate transmitter and receiver chains with their own antenna configurations. Hence, for each of the initiator and responder links, the transmitter and receiver can be trained independently to perform beamforming. The protocol used to train the transmitter during SLS is called TXSS, and the protocol used to train the receiver is called RXSS. This gives rise to four possibilities; if an ISS is used to the train the transmitter side of the initiator link, the phase is called ISS TXSS. Similarly, the other three possibilities are ISS RXSS, RSS TXSS, and RSS RXSS.

During TXSS, the transmitter sends a separate SSW frame from different available transmit sectors, the number of which can be pre-negotiated between stations. The receiver maintains a quasi-omni receive configuration. If the receiver has multiple antennas, the transmitter repeats this process for each receive antenna. The receiver measures the quality of the packet received from each of the transmit sectors by cycling through all of its receive antennas in quasi-omni mode. In the end, the receiver replies with the best sector. The standard allows a vendor-specific algorithm to decide the best sector. Similarly, during RXSS, the transmitter uses an omniantenna configuration, and the receiver changes receive sectors to determine which the best receive sector.

The SLS phase can be initiated by the PCP/AP during the BTI by performing an ISS. Then, the RSS and FB phases are completed during the A-BFT announced by the PCP/AP. In this case, the BRP phase completed in an ATI or DTI. Alternatively, a station can either use a CBAP period (also announced in the beacon) or schedule an SP to perform beamforming with another station. The DTI can be used for all the phases of beamforming (Fig. 11).

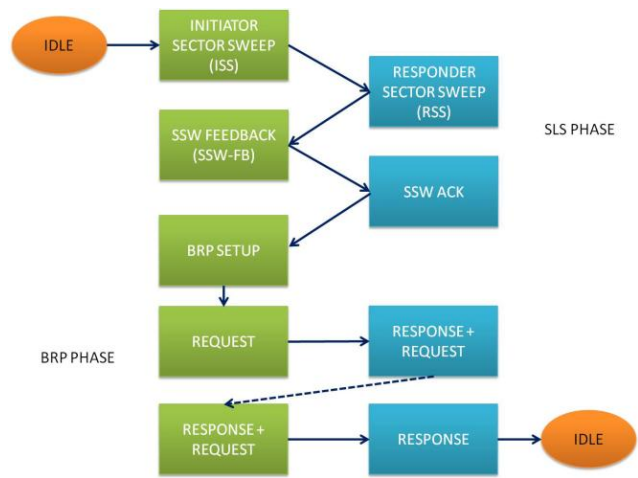

**Figure 13.** Parts of the sector level sweep and beam refinement phases of the beamforming in 802.11ad.
D Beam Refinement

The BRP phase comprises a setup phase followed by a beam-refinement phase based on request-response (Fig. 13). The request-response packets of the setup phase are exchanged until the responder (receiver) sets the capability-request field in the BRP packet at 0. This is followed by a response from the initiator (transmitter) with the capability-request field set at 0.

The beam-refinement request can be a transmitter- or receiver-refinement request. A transmitter-refinement request indicates the need for transmitter antenna training by the transmission station and vice versa. The transmitter station adds TRN-T subfields to the BRP frame. The receiver station holds data that it obtained by measuring the TRN-T fields. The receiver station responds to a receiver-refinement request (sent by the transmitter station) by appending TRN-R subfields to its response frames, that is, an ACK or block ACK frame.

The SLS and BRP phases of beamforming usually precede data transmission. They are completed right at the beginning of beamforming and are repeated periodically as needed. Beamtracking is used for beamforming training during data transmission to accommodate channel changes between two SLS/BRP beamforming training phases. In beamtracking, training fields comprising CE and STF fields are attached to the back of data packets or, for example, ACK/BA, to train the transmitter or receiver (Fig. 3). 802.11ad allows for three types of beamtracking, and the type of beamtracking is signaled using three parameters in the PLCP header. The three parameters of interest are packet type, training length, and beamtracking request. Of these, the training length is always greater than zero. If the training length is zero, the other two fields are reserved, and the packet does not contain any beamtracking training or request. Table 2 shows the types of training indicated by the PLCP parameters.

Table 2. Types of beamtracking indicated by the PLCP parameters.

| Parameters | BT type | Explanation |
| --- | --- | --- |
| BT request = 1<br>Packet Type = 1 | Send TRN-T | The transmitter attaches TRN-T fields to the current packet and the receiver sends back a BRP frame with the feedback. |
| BT request = 0<br>Packet Type = 0 | Send TRN-R | The transmitter attaches TRN-R fields to the current packet and the receiver finds its own best sector. |
| BT request = 1<br>Packet Type = 0 | Request TRN-R | The transmitter is requesting receiver training. In the next packet, the receiver attaches TRN-R fields. |

## 6 Power-Save Protocol

Dedicated SPs in 802.11ad allow battery-powered stations to hibernate during data transmission periods that are not assigned to them. An 802.11ad station can be in one of the two power-save states: doze or awake. When awake, a station is fully powered; when in doze, the station is powered off. A station's power-save state in various sections in a beacon interval depends on whether the station is in active mode or power-save mode. In power-save mode, stations with or without PCP/AP can doze for one or more consecutive BIs, or sections of a BI, more if they were permanently in active mode. A station must check its peer's wakeup schedule before sending any individually addressed MPDUs to the peer station because it may be in doze mode. A station without PCP/AP can always use information request/response frames to request the wakeup schedule from any of its peers if required. Fig. 14 shows power management modes and state transitions.

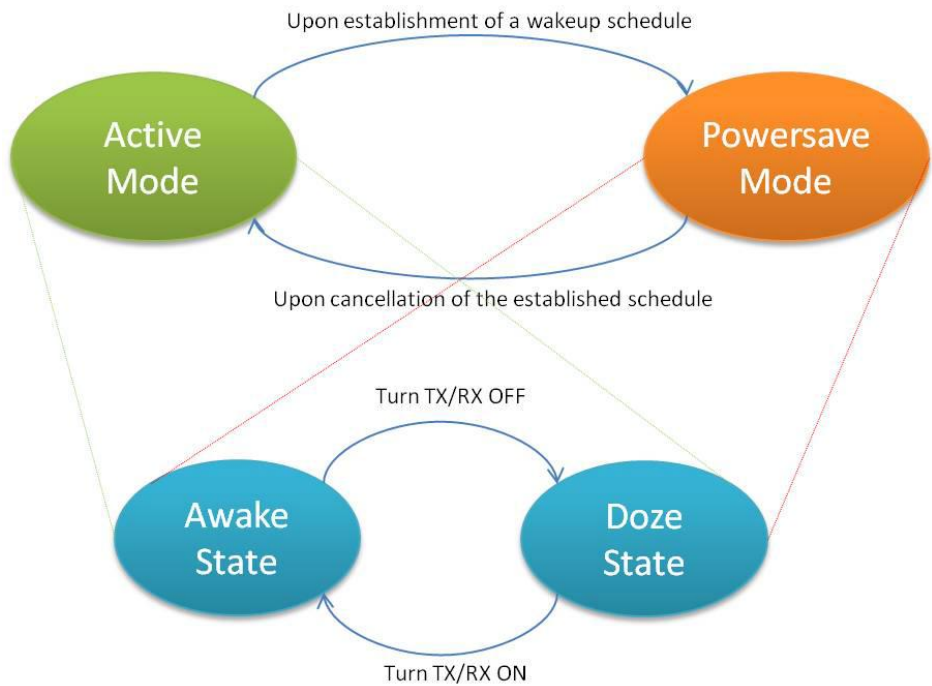

**Figure 14.** Power management-mode/state transitions.

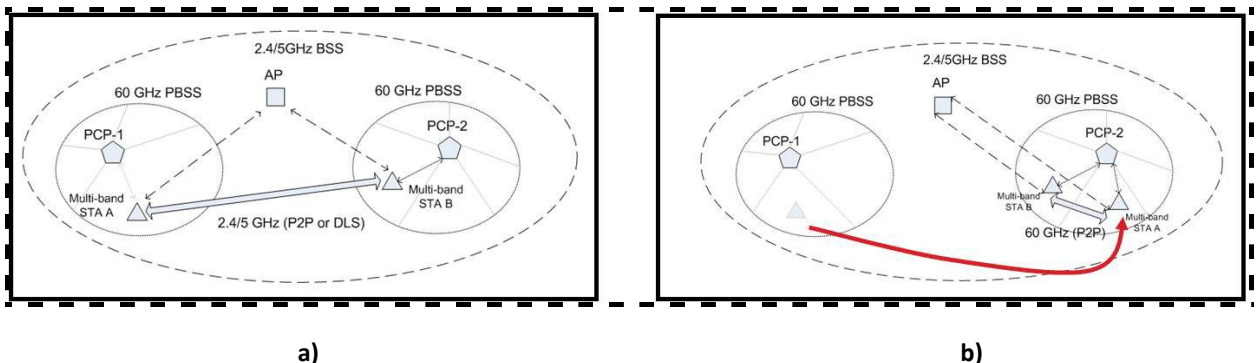

**Figure 15.** Fast session transfer done by peer stations (STAs) A and B in an a) 2.4 GHz channel and b) a 60 GHz channel.

A station without PCP/AP that has not established a wakeup schedule with its peer is in active mode. To switch from active to power-save mode, the station establishes a wakeup schedule with PCP/AP. It does this by including a wakeup schedule (WS) element in its power-save configuration request. To switch from power-save mode to active mode, the station without PCP/AP sends a power-save configuration request in which the power management bit is set at 0. The station immediately switches to active mode upon receipt of the ACK frame from PCP/AP.

A PCP/AP station includes its WS in its beacon or announcement frames before switching to power-save mode. When switching back to active mode, it ceases including the WS in these frames. The PCP/AP station keeps track of the wakeup schedules of all associated stations without PCP/AP. In addition, APs also have to buffer MPDUs addressed to associated stations in doze state and forward these MPDUs at designated times.

**7 Fast Session Transfer Protocol**

Fast session transfer (FST) protocol allows different streams or sessions to transfer smoothly from one channel to another in the same band or different bands. This protocol makes 802.11ad compatible with the forthcoming 802.11ac standard and other existing standards, such as 802.11a/b/g/n. The protocol allows different radios in the same device to operate simultaneously or not simultaneously. Devices with 802.11ac and 802.11ad can have same MAC address or different MAC addresses. If the same MAC address is used for all the radios in different bands then FST is in transparent mode. If the MAC addresses differ according to channel/band, then FST is not transparent.

A simple example of FST is a video stream to be established between STA A and STA B in the 2.4 GHz band using direct-link setup (DLS). The STAs are 40 m apart. The video uses 802.11n radio at 144.4 Mbit/s and H.264 compression. After some time, the user of STA A moves very close to STA B so that the separation is less than 3 m. Both STA A and STA B understand that they have 60 GHz radio, which was discovered during in 60 GHz discovery mode. They then transition to 60 GHz channel 2 and use an uncompressed stream by closing their link at MCS 12, which is 4.62 Gbit/s (Fig. 15).

This video stream established in 60 GHz channel 2 can be moved to 60 GHz channel 1 if there is congestion in channel 2. It can also be moved to channel 4 in 5 GHz or channel 6 in 2.4 GHz when STA A starts to move away from STA B. In this example, video compression, such as H.264 and that used in the WiGig WDE specification, ensures that the session does not drop because of large range or insufficient bandwidth. The application and MAC layers interact with PHY to optimize the smooth delivery of content to the end application. They compress the video whenever the band is not 60 GHz by using IEEE 802.11ac or IEEE 802.11n and then transition to uncompressed video using 60 GHz SC or OFDM modes when the range is less than three meters. This ensures the highest QoS. The FST also ensures that a subset of streams can be transferred from one channel/band to another while the remaining streams are in the original channel/band.

Fig. 16 shows the basic FST protocol. The initiator and responder are assumed to have one station management entity (SME) and two MAC layer management entities (MLMEs) that correspond to two different radios. In device discovery, both the STA and PCP recognize that they have multiple radios through beacon/information request and response frames. Then, the STA decides to establish an FST session with the PCP/AP in the other band or new channel. Any signal exchanges before the FST decision box is to establish the FST session with PCP/AP and is done in the current channel. Once the decision to transfer the session or stream has been decided at the predetermined times at both the STA and PCP/AP, the actual transition to the new channel occurs. Then, the basic FST ACK information is sent and exchanged to signal that both the STA and PCP/AP have transitioned to the new channel.

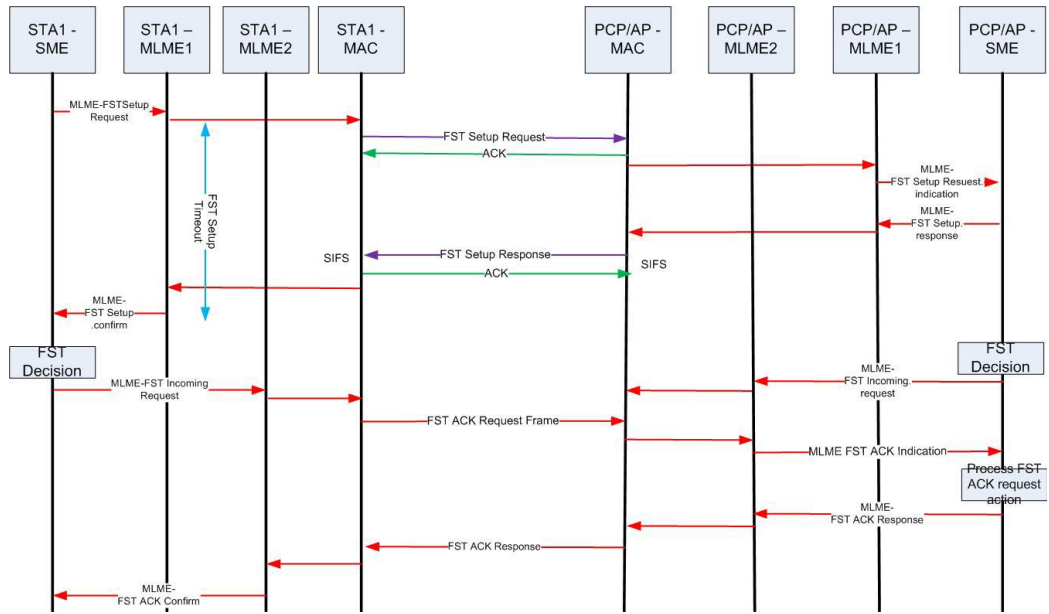

FST: fast session transfer
MLME: MAC layer management entities
PCP: PBSS control point
SME: Station management Entity
STA: peer station

**Figure 16.** FST protocol exchange between STA and PCP when moving from one band/channel to another.

## 8 Packet Throughput

The modulation and coding schemes in section 3 along with the normal ACK, A-MSDU, and A-MPDU packet structures in section 4 allow the 60 GHz radios to change between very different achievable throughputs depending on the packet size. Fig. 17 shows the MAC layer throughput versus packet size sent at different MCS values. Similarly, Fig. 18 shows the throughput versus packet size when A-MPDU is used. In Figs. 17 and 18, BTI, ABFT and AT overheads are not taken into account. The parameters described in section 3 can also be found in section 21.3 of [9].
All the throughput curves saturate with respect to the packet size. In general, longer packets can go through the channel without error if the channel remains constant for the entire duration; that is, longer packets need longer channel coherence time. Throughput curves saturate for larger packet sizes, and even though the channel might be good enough to support longer packets, it is better to use a higher MCS than bigger packets (through aggregation), once the length is close to the saturation point.

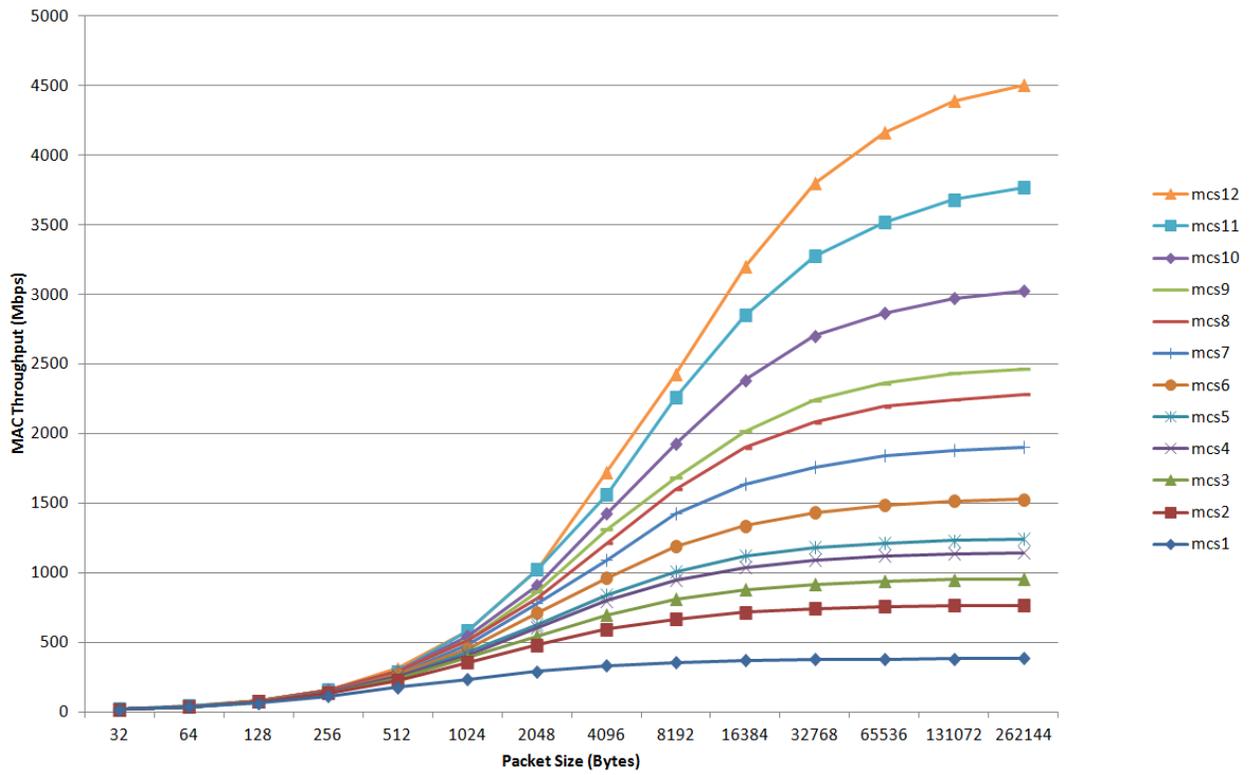

**Figure 17.** Single carrier throughput as a function of the packet size for different MCS values for non-aggregated packets.

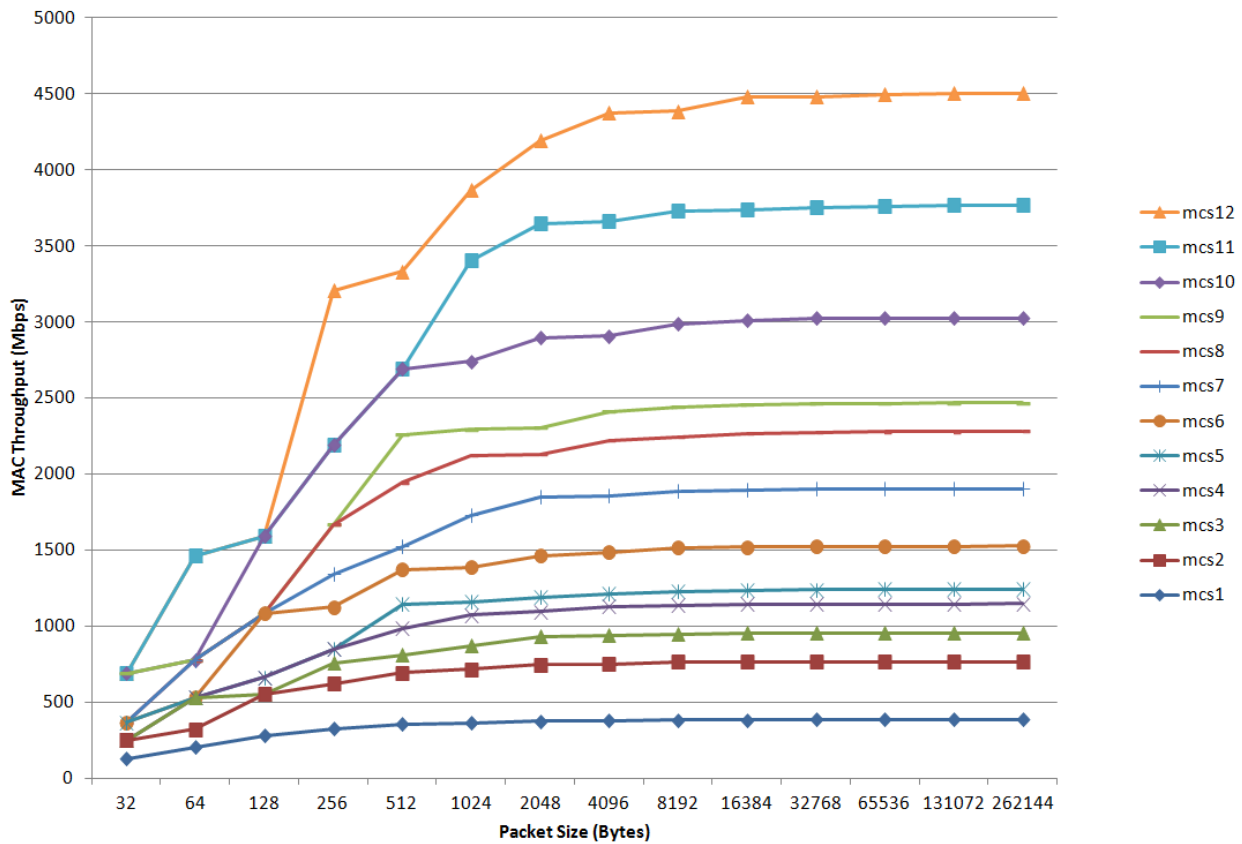

**Figure 18.** Single carrier throughput as a function of the packet size for different MCS values for A-MPDU packets.

## 9 Conclusion

802.11ad are standardizing 60 GHz technology to facilitate multi-gigabit-per-second communications over shorter distances. This standard has many new features to improve and sustain high-speed communications with TDMA single-carrier and OFDM schemes. They allow for scheduled and contention-based access, beamforming, and power-save mechanisms that decrease power consumption and increase throughput. Future evolution of 802.11ad towards full MIMO support and channel bonding can further increase its data rate. With the advent of new technologies to make these protocols practical, and with standardization by bodies such as WiGig and IEEE, truly wireless broadband will be achieved with 60 GHz, and all wires in PANs will be eliminated.

http://www.ieee802.org/11/Reports/cmmw update.htm

**Biographies**

**Sai Shankar N** received his Ph.D from Indian Institute of Science, Bangalore in the year 1998 and worked on DAAD fellowship on Queueing approaches in manufacturing at Dept. of Mathematics, University of Kaiserslautern, Germany. In 1999, he joined Philips Research, Eindhoven, Netherlands to work on IEEE 802.15 HFC networks and on Differentiated services. In 2001 he was transferred to Philips Research, New York and worked on IEEE 802.11e, IEEE 802.11n, MBOA UWB and IEEE 802.22. To this end he was nominated as one of the five finalists in EE Times for his contributions to UWB MAC. In 2005, he worked at Qualcomm in San Diego on IEEE 802.11s and RLC and MAC-hs issues in HSPA. In 2007 he worked at Broadcom Corporation working on 802.11 AMP in Bluetooth SIG and 60 GHz. Currently he is with Tensorcom leading its 60 GHz FW and MAC HW solution. He can be reached at nsai@tensorcom.com.

**Debashis Dash** received his B. Tech. degree from the Indian Institute of Technology, Kanpur, India, in 2004 and his M.S. degree from Rice University, Houston, in 2007. He is a Ph.D. candidate with the Department of Electrical and Computer Engineering, Rice University, Houston, TX. He currently works at Tensorcom, San Diego, CA. His research interests include information theory and graph theory and their applications in wireless systems.

**Hassan El Madi** received his B.S. degree in computer engineering from the University of California, San Diego in 2007 and his M.Eng. degree in electrical engineering with an emphasis in wireless communications from Virginia Tech, Blacksburg, VA in 2010. He currently works as a software staff engineer at Tensorcom, San Diego, CA. His interests include cognitive radios (spectrum sensing, automatic modulation classification, geolocalization) as well as the design and implementation of WLAN and WPAN MAC and PHY layers.

**Guru Gopalakrishnan** received his B.E. degree in Electronics and Communication from Anna University, India in 2006 and his M.S. degree in Electrical Engineering (Computer Networks) from University of Southern California, Los Angeles in 2009. He earlier worked at Broadcom Corporation, San Diego in Bluetooth and Bluetooth low energy technologies and is currently at Adeptence, San Diego. His research interests include throughput and power optimizations for wireless systems.